\documentclass[aps,preprint,amssymb,12pt,floatfix]{revtex4}
\usepackage{subfigure}
\usepackage{graphicx}
\usepackage{rotating} \usepackage{color}
 \usepackage{amsmath,amsthm}
\usepackage{epstopdf} 
 \def\rv{{\mathbf{r}}}
\def\fv{{\mathbf{f}}} 

\begin{document}
\title{Force Dependent Hopping Rates of RNA Hairpins can be Estimated from Accurate Measurement of the Folding Landscapes}
\author{Changbong Hyeon$^{1\dagger}$, Greg Morrison$^{2,3\dagger}$ and D. Thirumalai$^{2,4}$}
\thanks{Corresponding author phone: 301-405-4803; fax: 301-314-9404; thirum@umd.edu}
\affiliation{
$^1$Department of Chemistry, Chung-Ang University, Seoul 156-756, Republic of Korea\\
$^2$Biophysics Program, Institute For Physical Science and Technology, University of Maryland, College Park, Maryland 20742\\
$^3$Department of Physics, University of Maryland, College Park, Maryland 20742\\
$^4$Department of Chemistry, University of Maryland, College Park, Maryland 20742\\
$^{\dagger}$C.H. and G.M. contributed equally to this work.}  
\date{\today}

\begin{abstract}
The sequence-dependent folding landscapes of nucleic acid hairpins reflect much of the complexity of biomolecular folding.  Folding trajectories, generated using single molecule force clamp experiments by attaching semiflexible polymers to the ends of hairpins have been used to infer their folding landscapes.  Using simulations and theory, we study the effect of the dynamics of the attached handles on the handle-free RNA free energy profile $F^o_{eq}(z_m)$, where $z_m$ is the molecular extension of the hairpin.  Accurate measurements of $F^o_{eq}(z_m)$ requires stiff polymers with small $L/l_p$, where $L$ is the contour length of the handle, and $l_p$ is the persistence length.  Paradoxically, reliable estimates of the hopping rates can only be made using flexible handles.  Nevertheless, we show that the equilibrium free energy profile $F^o_{eq}(z_m)$ at an external tension $f_m$, the force ($f$) at which the folded and unfolded states are equally populated, in conjunction with Kramers' theory, can provide accurate estimates of the force-dependent hopping rates in the absence of handles at arbitrary values of $f$.  Our theoretical framework shows that $z_m$ is a good reaction coordinate for nucleic acid hairpins under tension.
\end{abstract}

\maketitle

A molecular understanding of how proteins and RNA fold is needed to describe the functions of enzymes \cite{FershtBook} and ribozymes \cite{DoudnaNature02}, interactions between biomolecules, and the origins of misfolding that is linked to a number of diseases \cite{Dobson99TBS}.  The energy landscape perspective has provided a conceptual framework for describing the mechanisms by which unfolded molecules navigate the large conformational space in search of the native state \cite{DillNSB97,OnuchicCOSB04,HyeonBC05}.  Recently, single molecule techniques have been used to probe features of the energy landscape of proteins and RNA that are not easily accessible in ensemble experiments \cite{Fisher00NSB,Bustamante01Science,Haran03PNAS,TinocoBJ06,SchulerNATURE2002,ZhuangCOSB03,EvansNature99,Woodside06PNAS,Block06Science,Li07PNAS,DietzPNAS04,Mickler07PNAS}.
It is possible to construct the shape of the energy landscape, including the energy scales of ruggedness \cite{HyeonPNAS03,ReichEMBOrep05}, using dynamical trajectories that are generated by applying a constant force ($f$) to the ends of proteins and RNA \cite{Schlierf04PNAS,Woodside06PNAS,Fernandez06NaturePhysics,Block06Science}.  If the observation time is long enough for the molecule to sample the accessible conformational space, then the time average of an observable $X$ recorded for the $\alpha^{th}$ molecule ($\langle X\rangle=\lim_{t\rightarrow\infty}\frac{1}{t}\int^t_0d\tau X_{\alpha}(\tau)$) should equal the ensemble average ($\langle X\rangle=\lim_{N\rightarrow \infty}\frac{1}{N}\sum_{i=1}^NX_i$), and the distribution $P(X)$ should converge to the equilibrium distribution function $P_{eq}(X)$.
Using this strategy, laser optical tweezer (LOT) experiments have been used to obtain the sequence-dependent folding landscape of a number of RNA and DNA hairpins \cite{Bustamante01Science,Woodside06PNAS,Bustamante03Science,Block06Science}, using $X=R_m$, the end-to-end distance of the hairpin that is conjugate to $f$, as a natural reaction coordinate.
In LOT experiments, the hairpin is held between two long handles (DNA \cite{Block06Science} or DNA/RNA hybrids \cite{Bustamante01Science}), whose ends are attached to polystyrene beads (Fig. 1a).  The equilibrium free energy profile $\beta F_{eq}(R_m)=-\log{P_{eq}(R_m)}$ ($\beta\equiv1/k_BT$, $k_B$ is the Boltzmann constant, and $T$ is the absolute temperature) may be useful in describing the dynamics of the molecule, provided $R_m$ is an appropriate reaction coordinate.

The dynamics of the RNA extension in the presence of $f$ ($z_m=z_{3'}-z_{5'}\approx R_m$, provided transverse fluctuations are small) is indirectly obtained in an LOT experiment by monitoring the distance between the attached polystyrene beads ($z_{sys}=z_p-z_o$), one of which is optically trapped at the center of the laser focus (Fig 1a).  The goal of these experiments is to extract the folding landscape ($\beta F_{eq}^o(z_m)$) and the dynamics of the hairpin in the absence of handles, using the $f$-dependent trajectories $z_{sys}(t)$.  To achieve these goals, the fluctuations in the handles should minimally perturb the dynamics of the hairpin in order to probe the true dynamics of a molecule of interest.  
However, depending on $L$ and $l_p$ ($L$ is the contour length of the handle and $l_p$ is its persistence length), the intrinsic fluctuations of the handles can not only disort the signal from the hairpin, but also directly affect its dynamics.
The first is a problem that pertains to the measurement process, while the second is a problem of the coupling between the instruments and the dynamics of RNA.

Here, we use coarse-grained molecular simulations of RNA hairpin and theory to show that, in order to obtain accurate $\beta F_{eq}(R_m)$, the linkers used in the LOT have to be stiff, i.e., the value of $L/l_p$ has to be small.  To investigate the handle effects on the energy landscape and hopping kinetics, we simulated the
hairpin dynamics under force-clamp conditions by explicitly modeling the linkers as polymers with varying $L$ and $l_p$.  Surprisingly, the force-dependent folding and unfolding rates that are directly measured using the time traces, $z_m(t)$, are close to the ideal values (those that are obtained by directly applying $f$, without the handles, to the 3' end with a fixed 5' end) only when the handles are flexible.  Most importantly,  accurate estimates of the $f$-dependent hopping rates over a wide range of $f$-values, {\em{in the absence of handles}}, can be made using $\beta F_{eq}(R)$, in the presence of handles obtained at $f=f_m$, the transition midpoint at which the native basin of attraction (NBA) and the unfolded basin of attraction (UBA) of the RNA are equally populated.  The physics of a hairpin attached to handles is captured using a generalized Rouse model (GRM), in which there is a favorable interaction between the two non-covalently linked ends.  The GRM gives quantitative agreement with the simulation results.  The key results announced here provide a framework for using the measured folding landscape of nucleic acid hairpins at $f\approx f_m$ to obtain $f$-dependent folding and unfolding times and the transition state movements as $f$ is varied \cite{KlimovPNAS99,HyeonPNAS05,RitortPRL06,West06BJ,Dudko06PRL,Hyeon07JP}.\\

\bigskip
{\bf RESULTS and DISCUSSIONS}
\bigskip

{\bf Modeling the LOT experiments:}  
In order to extract the folding landscape from LOT experiments, the time scales associated with the dynamics of the beads, handles, and the hairpin have to be well-separated \cite{RitortBJ05,HyeonBJ06,Manosas07BJ,WestPRE}. 
The bead fluctuations are described by the overdamped Langevin equation $\gamma dx_p/dt=-kx_p+F(t)$ where $k$ is the spring constant associated with the restoring force, and the random white-noise force $F(t)$ satisfies $\langle F(t)\rangle=0$ and $\langle F(t)F(t')\rangle=2\gamma k_BT\delta(t-t')$.
The bead relaxes to its equilibrium position on a time scale $\tau_r=\gamma/k$. In terms of the trap stiffness, $k_p$, and the stiffness $k_m$ associated with the Handle-RNA-Handle (H-RNA-H; see Fig. 1), $k=k_p+k_m$.  With $\gamma=6\pi\eta a$, $a=1\mu$m, $\eta\approx 1$cP, $k_p\approx 0.01$ pN/nm \cite{Block94ARBBS}, and $k_m\approx 0.1$ pN/nm, we find $\tau_r\lesssim 1$ ms.  
In LOT experiments \cite{RitortBJ05,Manosas07BJ,WestPRE}, separation in time scales is satisfied such that $\tau^o_{U}\approx \tau_F^o\gg\tau_r$ at $f\approx f_m$, where $\tau_U^o$ and $\tau_F^o$ are the intrinsic values of the RNA (un)folding times in the absence of handles.

Since $z_m$ is a natural reaction coordinate in force experiments, the dispersion of the bead position may affect the measurement of $F_{eq}(z_m)$.  At equilibrium, the fluctuations in the bead positions satisfy $\delta x^2_{eq}\sim k_BT/(k_p+k_m)\sim k_BT/k_m$, and hence $k_m$ should be large enough to minimize the dispersion of the bead position.  The force fluctuation, $\delta f^2_{eq}\sim k_BTk_p^2/(k_p+k_m)$, is negligible in the LOT because $k_p\ll k_m$, and as a result $\delta f_{eq}/f_m\sim 0$, since $\delta f_{eq}\approx 0.1$ pN while $f_m\sim 15$ pN.  Thus, we model the LOT setup by assuming that the force and position fluctuations due to the bead are small, and exclusively focus on the effect of handle dynamics on the folding landscape and hopping kinetics of RNA (Figs. 1a-b).  \\

{\bf Short, stiff handles are required for accurate free energy profiles}:  For purposes of illustration, we used the self-organized polymer (SOP) model of the P5GA hairpin \cite{HyeonBJ07}, and applied a force $f=f_m\approx 15.4$ pN.  The force is exerted on the end of the handle attached to the 3' end of the RNA (P in Fig.1a), while fixing the other end (O in Fig.1a).  Simulations of P5GA  with handles of length $L=25$ nm and persistence length $l_p=70$ nm show that the extension of the entire system ($z_{sys}=z_p-z_0$) fluctuates between two limits centered around $z_{sys}\approx 50$ nm and $z_{sys}\approx 56$ nm (Fig. 1b).  
The time-dependent transitions in $z_{sys}$ between 50 nm and 56 nm correspond to the hopping of the RNA between the Native Basin of Attraction (NBA) and Unfolded Basin of Attraction (UBA).  
Decomposition of $z_{sys}$ as $z_{sys}=z_H^{5'}+z_m+z_H^{3'}$, where $z_H^{5'}(=z_{5'}-z_o)$ and $z_H^{3'}(=z_p-z_{3'})$ are the extensions of the handles parallel to the force direction (Fig. 1a), shows that $z_{sys}(t)$ reflects the transitions in $z_m(t)$ (Fig. 1b).  
Because the simulation time is long enough for the harpin to ergodically explore the conformations between the NBA and UBA, the histograms collected from the time traces amount to the equilibrium distributions $P_{eq}(X) $ where $X=z_{sys}$, $z_H^{5'}$, $z_m$, or $z_H^{3'}$ (Fig 1b; for $P_{eq}(z_H^{5'})$ and $P_{eq}(z_H^{3'})$, see the Supporting Information SI Fig. 6a).  To establish that the time traces are ergodic, we show that $\overline{z}_T(t)=\frac{1}{t}\int^t_0d\tau z_{sys}(\tau)$ reaches the thermodynamic average ($\approx$ {$\int_{-\infty}^{\infty}z_{sys}P_{eq}(z_{sys})dz_{sys}$=53.7 nm}) after $t\gtrsim 0.1$ sec (the magenta line on $z_{sys}(t)$ in Fig. 1b).

Fig. 1b shows that the positions of the handles along the $f$ direction fluctuate, even in the presence of tension, which results in slight differences between $P_{eq}(z_{sys})$ and $P_{eq}(z_m)$.  Comparison between the free energy profiles obtained from the $z_{sys}(t)$ and $z_m(t)$ can be used to investigate the effect of the characteristics of the handles on the free energy landscape.  
To this end, we repeated the force-clamp simulations by varying the contour length ($L=5-100$ nm) and persistence length ($l_p=0.6$ and 70 nm) of the handles.  
Fig. 2 shows that the discrepancy between the measured free energy $F_{eq}(z_{sys})$ (dashed lines in blue) and the molecular free energy $F_{eq}(z_m)$ (solid lines in red) increases for the more \emph{flexible} and \emph{longer} handles (see the SI text and SI Fig. 6 for further discussion of the dependence of the handle fluctuations on $L$ and $l_p$).  For small $l_p$ and large $L$, the basins of attraction in $F_{eq}(z_m)$ are not well resolved.  The largest deviation between $F_{eq}(z_{sys})$ and $F_{eq}(z_m)$ is found when $l_p=0.6$ nm and $L=25$ nm ($L/l_p\approx 40$) (the graph enclosed by the orange box in Fig. 2a). 
In contrast, the best agreement between $F_{eq}(z_{sys})$ and $F_{eq}(z_m)$ is found for $l_p=70$ nm and $L=5$ nm (the graph inside the magenta box in Fig. 2), which corresponds to $L/l_p\approx 0.07$. 
In the LOT experiments, $L/l_p\approx 6-7$ \cite{Bustamante01Science,Woodside06PNAS,Block06Science}. \\
 
{\bf{Generalized Rouse model (GRM) captures the physics of H-RNA-H under tension}}:  In order to establish the generality of the relationship between the free energy profiles as measured by $z_m$ and those measured by $z_{sys}$, we introduce an exactly solvable model that minimally represents the RNA and handles (Fig. 3a).  We mimic the hairpin using a Gaussian chain with $N_{0}$ monomers and Kuhn length $a$.  The endpoints of the RNA mimic are harmonically trapped in a potential with stiffness $k$ as long as they are within a cutoff distance $c=4$nm.  Two handles, each with $N_h$ monomers and Kuhn length $b$, are attached to the ends of the RNA (see Methods). We fix one endpoint of the entire chain at the origin, and apply a force $f_m\approx15.4$ pN to the other end.  
The free energies as a function of both the RNA's extension, $R_m=|\mathbf{r}_{3'}-\mathbf{r}_{5'}|$ ($\approx z_m$ at high $f$) and the system's extension $R_{sys}=|\rv_{P}-\rv_0|$ ($\approx z_{sys}$ at high $f$) are exactly solvable in the continuum representation.  We choose $k$ such that $f_m$ is near the midpoint of the transition, so that $\int_0^c d^3\rv\,P_{eq}(\rv)\approx \int_c^\infty d^3\rv\,P_{eq}(\rv)$. We tune $N_0$ so that the barrier heights for the GRM and P5GA are similar at $f=f_m$.  These requirements give $N_0= 20$ and $k\approx 0.54$ pN/nm. 

While the stiffness in the handles of the simulated system (Fig. 1) cannot be accurately modeled using a Gaussian chain, the primary effect of attaching the handles is to alter the fluctuations of the endpoints of the RNA.  By equating the longitudinal fluctuations for the WLC, $\langle\delta{\mathbf{R}}^2_{||}\rangle_{WLC}\sim Ll_p^{-1/2}(\beta f)^{-3/2}$, with the fluctuations for the Gaussian handles, $\langle\delta{\mathbf{R}}^2_{||}\rangle_{G}\sim Lb$, we  estimate that the effective persistence length of the handles scales as $l_p^{eff}\sim b^{-2} f^{-3}$ (see the SI for details).  
Thus, smaller spacing in the Gaussian handles in the GRM will mimic stiffer handles in the H-RNA-H system.  The free energies computed for the GRM, shown in Fig. 3b-c, are consistent with the results of the simulations.  The free energy profiles deviate significantly from $F^o_{eq}(z_m)$ as $N_h$ increases or `stiffness' decreases.  The relevant variable that determines the accuracy of  $F_{eq}(z_{sys})$ is $N_h b^2\sim L/l_p^{eff}$, with the free energies remaining unchanged if $N_hb^2$ is kept constant.  The barrier height and well depths as a function of $z_m$ are unchanged as a function of $L$ and $b$.  However, the apparent activation energy is decreased as measured by $z_{sys}$ (seen in Fig. 2 as well). The GRM confirms that  accurate measurement of the folding landscape using $z_{sys}$ requires stiff handles.\\

{\bf Accurate estimates of the hopping kinetics requires short and flexible handles:} 
Because LOT experiments can also be used to measure the force-dependent rates of hopping between the NBA and the UBA, it is important to assess the influence of the dynamics of the handles on the intrinsic hopping kinetics of the RNA hairpin.  In other words, how should the structural characteristics of the linkers be chosen so that the measured hopping rates using the time traces $z(t)$ and the intrinsic rates are as close as possible?

{\it Folding and unfolding rates of P5GA and the free energy profile without handles :} We first performed force clamp simulations of P5GA in the absence of handles to obtain the intrinsic (or ideal) folding ($\tau_F^o$) or unfolding ($\tau_U^o$) times, that serve as a reference for the H-RNA-H system.  To obtain the boundary conditions for calculating the mean refolding and unfolding times, we collected the histograms of the time traces and determined the positions of the minima of the NBA and UBA, $z_F=1.9$ nm and $z_U=7.4$ nm (Fig. 4a).  The analysis of the time traces provides the transition times in which $z_m$ reaches $z_m=z_U$ starting from $z_m=z_F$.  The mean unfolding time $\tau_U$ is obtained using either $\tau_U=1/N\sum_i\tau_U(i)$, or from the fits to the survival probability $P_F(t)=e^{-t/\tau_U}$ (SI Fig. 8). 
The mean folding time is similarly calculated, and the two methods give similar results.  The values of $\tau_U^o$ and $\tau_F^o$ computed from the time trace of $z_m(t)$ are 2.9 ms and 1.9 ms, respectively.  At $f_m=15.4$ pN and $L=0$ nm, the equilibrium constant $K_{eq}=\tau^o_F/\tau^o_U=0.67$,
which shows that the bare molecular free energy is slightly tilted towards NBA at $f=15.4$ pN.  

{\it Hopping times depend on the handle characteristics :} 
The values of the folding ($\tau_F^m$) and unfolding($\tau_U^m$) times were also calculated for the P5GA hairpin with attached handles (Fig. 1).  As the length of the handles increases {\it{both $\tau_U^m$ and $\tau_F^m$ increase}} gradually, and the equilibrium distribution shifts towards the UBA, i.e. $K_{eq}=\tau^m_F/\tau_U^m$ increases (Fig. 4b).   
Strikingly, the use of flexible handles results in minimal deviations of $\tau_U^m$ and $\tau_F^m$ from their intrinsic values (Fig.4b).  
Attachment of handles (stiff or flexible) to the 5' and 3' ends restricts their movement, which results in a decrease in the number of paths to the NBA and UBA. Thus, both $\tau_U^m$ and $\tau_F^m$ increase (Fig.4b). 
As the stiffness of the handle increases the extent of pinning increses.  These arguments show that \emph{flexible and short handles, that have the least restriction on the fluctuations of the 5' and 3' ends of the hairpin,  cause minimal perturbation to the intrinsic RNA dynamics}, and
hence the hopping rates. 

Because the experimentally accessible quantity is the extension of the H-RNA-H, 
it is important to show that the transition times can be reliably obtained using $z_{sys}(t)$.  Although $z_{sys}(t)$ differs from $z_m(t)$ in  amplitude, the ``phase'' between the two quantities track each other reliably throughout 
the simulation, even when the handles are long and flexible (see SI Fig.7).  
We calculated $\tau_U^{sys}$ and $\tau_F^{sys}$ by analyzing the trajectories $z_{sys}(t)$ using the same procedure used to compute their intrinsic values. 
Comparison of $\tau_U^{sys}$ ($\tau_F^{sys}$) and $\tau_U^m$ ($\tau_F^m$) for both stiff and flexible handles shows excellent agreement at all $L$ values 
(Fig.4b). 
Thus, it is possible to infer the RNA dynamics $z_m(t)$ by measuring $z_{sys}(t)$.

{\it{Theoretical predictions using the GRM are consistent with the simulations}}:  The 
simulation results can be fully understood using 
the GRM (Fig 3a), for which we can exactly solve the overdamped Langevin equation using the discrete representation of the Gaussian chain (see Methods).  
By assuming that transverse fluctuations are small (which is reasonable under the relatively high tension of $f=15.4$ pN), we use the Wilemski and Fixman (WF) theory \cite{FixmanJCP74II} to determine an approximate time of contact formation ($\tau_F^m=(k_F^m)^{-1}$) as a function of $b$ (i.e. increasing handle `stiffness') and $N_h$.   
The refolding rate of the RNA hairpin under tension is analogous to $k_F^m$.  A plot of $k^m_F(L)/k^m_F(0)$ versus $L$ (Fig. 4c) illustrates that smaller deviations from the handle-free values occur when $l_p$ is small.  
Moreover, Fig. 4c shows that the refolding rate decreases for increasing $N_h$ {\it{regardless of the stiffness of the chain}}.   
The saturating value of $k_F^m$ as $N_h\to\infty$ depends on $b$, with `stiffer' handles having a much larger effect on the folding rate.  While the handles used in LOT experiments are significantly longer than the handle lengths considered here, the saturation of the folding rate suggests that $L\sim 100$ nm is sufficiently long for finite-size effects to be negligible. 

We also find the dependence of $k_{F}$ on $L$ agrees well with the behavior observed in the simulation of P5GA.  
The ratio $k_F^m(L)/k_F^m(0)$ for $b=a$ agrees well with the trends of the flexible linker ($l_p=0.6$ nm) for all of the simulated lengths, with both saturating at $k_F(L)\approx 0.35 k_F(0)$ for large $L$.  
The trends for `stiffer' chains (smaller $b$) in the GRM qualitatively agree with the P5GA simulation with stiff handles ($l_p=70$ nm), with remarkably good agreement for $0.1\le b/a\le 0.2$ over the entire range of $L$.  
The GRM, which captures the physics of both the equilibrium and kinetic properties of the more complicated H-RNA-H, provides a theoretical basis for extracting kinetic information from experimentally (or computationally) determined folding landscapes.\\

{\bf Free energy landscapes and hopping rates:}
Stiff handles are needed to obtain $F_{eq}(z_{sys})$ \cite{Block06Science} that resembles $F_{eq}^o(z_m)$, whereas the flexible handles produce hopping rates that are close to their handle-free values. These two findings appear to demand two mutually exclusive requirements in the choice of the handles in LOT experiments. However, if $z_m$ is a good reaction coordinate, then it should be possible to extract the hopping rates using accurately measured $F_{eq}(z_{sys})(\approx F_{eq}(z_m)\approx F_{eq}^o(z_m))$ at $f\approx f_m$, using handles with small $L/l_p$.  The (un)folding times can be calculated using the mean first passage time (Kramers' rate expression) with appropriate boundary conditions \cite{ZwanzigBook}, 
\begin{eqnarray}
\tau^{KR}_{U}&=&\int^{z_U}_{z_F} dye^{\beta F_{eq}(y)}\frac{1}{D_U}\int^y_{z_{min}} dx e^{-\beta F_{eq}(x)},\nonumber\\
\tau^{KR}_{F}&=&\int^{z_U}_{z_F}dye^{\beta F_{eq}(y)}\frac{1}{D_F}\int^{z_{max}}_y dx e^{-\beta F_{eq}(x)},
\label{eqn:MFPT}
\end{eqnarray}
where $z_{min}$, $z_{max}$, $z_U$ and $z_F$ are defined in Fig. 4a.  The effective diffusion coefficient $D_F(D_U)$ is obtained by equating $\tau_F^{KR}$ ($\tau_U^{KR}$) in equation (\ref{eqn:MFPT}), with $F_{eq}(z_m)=F_{eq}^o(z_m)$, to the simulated 
$\tau_F^o\ (\tau_U^o)$.   We calculated the $f$-dependent $\tau_U^m(f)$ and $\tau_F^m(f)$ by evaluating equation (\ref{eqn:MFPT}) using $F_{eq}^o(z_m|f)=F_{eq}^o(z_m|f_m)-(f-f_m)\cdot z_m$.   The calculated and simulated results for P5GA are in good agreement (Fig 5a-b). At the higher force ($f=16.8$ pN), the statistics of hopping transition within our simulation time is not sufficient to establish ergodicity. As a result, the simulation results are not as accurate at high forces (see SI Fig. 9).  To further show that the use of $F_{eq}^o(z_m|f)$ in equation (\ref{eqn:MFPT}) gives accurate hopping rates, we calculated $\tau_U^o(f)$ for the GRM and compared the results with direct simulations of the handle-free GRM, which allows the study of a wider range of forces (see Methods).  The results in Fig. 5c show that $F_{eq}^o(z_m|f)$ indeed gives very accurate values for the transition times from the UBA and NBA over a wide force range.  

\section*{CONCLUSIONS}

The self-assembly of RNA and proteins may be viewed as a diffusive process in a multi-dimensional folding landscape.  To translate this physical picture into a predictive tool, it is important to discern a suitable low-dimensional representation of the complex energy landscape, from which the folding kinetics can be extracted.   
Our results show that, in the context of nucleic acid hairpins, precise measurement of the sequence-dependent folding landscape of RNA is sufficient to obtain good estimates of the $f$-dependent hopping rates in the absence of handles.  It suffices to measure $F_{eq}(z_{sys})\approx F_{eq}(z_m)\approx F_{eq}^o(z_m)$ at $f=f_m$ using stiff handles, while $F_{eq}(z_m|f)$ at other values for $f$ can be obtained by tilting $F_{eq}(z_m|f_m)$.  The accurate computation of the hopping rates using $F_{eq}(z_m)$ show that $z_m$ is an excellent reaction coordinate for nucleic acid hairpins under tension.  Further theoretical and experimental work is needed to test if the proposed framework can be used to predict the force dependent hopping rates for other RNA molecules that fold and unfold through populated intermediates.  
\\ 

\section*{METHODS}

{\it{RNA hairpin:  }}The Hamiltonian for the RNA hairpin with $N$ nucleotides, which is modeled using the self-organized polymer (SOP) model \cite{HyeonBJ07}, is\begin{eqnarray}
H_{SOP}&=&-\frac{k R_0^2}{2}\sum_{i=1}^{N-1}\log\bigg(1-\frac{(r_{i,i+1}-r_{i,i+1}^o)^2}{R_0^2}\bigg)+\sum_{i=1}^{N-3}\sum_{j=i+3}^N\epsilon_h\bigg[\bigg(\frac{r_{i,j}^o}{r_{i,j}}\bigg)^{12}-2\bigg(\frac{r_{i,j}^o}{r_{i,j}}\bigg)^6\ \bigg]\Delta_{i,j}\nonumber\\
&&\qquad\qquad\qquad +\sum_{i=1}^{N-3}\sum_{j=i+3}^{N}\epsilon_l\bigg(\frac{\sigma}{r_{i,j}}\bigg)^{12}(1-\Delta_{i,j})+\sum_{i=1}^{N-2}\epsilon_l\bigg(\frac{\sigma^*}{r_{i,i+2}}\bigg)^6,\label{RNAHam}
\end{eqnarray}
where $r_{i,j}=|\rv_i-\rv_j|$ and $r^o_{i,j}$ is the distance between monomers $i$ and $j$ in the native structure.  The first term enforces backbone chain connectivity using the finite extensible nonlinear elastic (FENE) potential, with $k\approx 1.4\times 10^4$ pN$\cdot$nm$^{-1}$ and $R_0=0.2$ nm.  The Lennard-Jones interaction (second term in equation (\ref{RNAHam})) describes interactions only between native contacts (defined as $r^o_{i,j}\le 1.4$ nm for $|i-j|>2$), with $\Delta_{i,j}=1$ if monomers $i$ and $j$ are within 1.4 nm in the native state, and $\Delta_{i,j}=0$ otherwise.  
Non-native interactions are treated as purely repulsive (the third term in equation (\ref{RNAHam})) with $\sigma=0.7$ nm.  We take $\epsilon_h=4.9$ pN$\cdot$nm and $\epsilon_l=7.0$ pN$\cdot$nm for the strength of interactions. In the fourth term, the repulsion between the $i^{th}$ and $(i+2)^{th}$ interaction sites along the backbone has $\sigma^*=0.35$ nm to prevent disruption of the native helical structure.

{\it{Handle polymers:  }}The handles are modeled using the Hamiltonian
\begin{equation}
H_{handles}=\frac{k_S}{2}\sum^{N-1}_{i=1}({{r}}_{i,i+1}-r_0)^2-k_A\sum^{N-2}_{i=1}\hat{\mathbf{r}}_{i,i+1}\cdot\hat{\mathbf{r}}_{i+1,i+2}\label{HandleHam}. 
\end{equation}
The neighboring interaction sites, with an equilibrium distance $r_0=0.5$ nm, are harmonically constrained with a spring constant $k_S\approx 1.4\times 10^4$ pN$\cdot$nm$^{-1}$.  
In the second term of eq. (\ref{HandleHam}), the strength of the bending potential, $k_A$, determines the handle flexibility. 
We choose two values, $k_A=$7.0 pN$\cdot$nm and $k_A=$561 pN$\cdot$nm to model flexible and semiflexible handles respectively, and assign $k_A=35$ pN$\cdot$nm to the junction connecting two ends of the RNA and the handles.  
We determine the corresponding persistence length for the two $k_A$ values as $l_p=0.6$ and 70 nm (see SI text). 
The contour length of each handle is varied from $N=5-200$.  
\\
{\it{Generalized Rouse model (GRM):}} The Hamiltonian for the GRM (Fig. 3a) is
\begin{eqnarray}
\beta H&=&\frac{3}{2b^2}\sum_{i=1}^{N_h} (\rv_{i+1}-\rv_i)^2+\frac{3}{2b^2}\sum_{i=N_h+N_0+1}^{2N_h+N_0} (\rv_{i+1}-\rv_i)^2-\beta\fv\cdot(\rv_N-\rv_1)+\beta k_0 \rv_1^2\nonumber\\
&&\qquad\qquad\qquad+\frac{3}{2a^2}\sum_{i=N_h+1}^{N_h+N_0} (\rv_{i+1}-\rv_i)^2+\beta V[\rv_{N-N_h+1}-\rv_{N_h+1}],\label{TheoryHam}
\end{eqnarray}
where
\begin{eqnarray}
V[\rv]=\left\{\begin{array}{cc} k \rv^2 & |\rv|\le c \\kc^2 & |\rv|>c \end{array}\right. .\label{CutPot}
\end{eqnarray}
The first two terms in equation (\ref{TheoryHam}) are the discrete connectivity potentials for the two handles, each with $N_h$ bonds ($N_h+1$ monomers), and with Kuhn length $b$. The mechanical force $\fv$ in the third term is applied along the $z$ direction, with $|\fv|=f_m=15.4$ pN.  We also fix the 5' end of the system with a harmonic bond of strength $k_0=2.5\times 10^4$ pN$\cdot$nm$^{-1}$ in the fourth term of eq. (\ref{TheoryHam}).  The fifth term mimics the RNA hairpin with $N_0$ bonds and spacing $a=0.5$ nm.  Interactions between the two ends of the RNA hairpin are modeled as harmonic bond with strength $k\approx 0.54$ pN$\cdot$nm$^{-1}$ that is cut off at $c=4$ nm (eq. (\ref{CutPot})).  When $|{\mathbf{R}}_m|$ exceeds 4nm, the bond is broken, mimicking the unfolded state.

The free energies as a function of both $R_m\approx z_m$ and $R_{sys}\approx z_{sys}$ are most easily determined in the continuum limit of the Hamiltonian in equation (\ref{TheoryHam}), with $\sum_{i=1}^N\to \int_0^N ds$.  Because of the relatively large value of the external tension ($f_m\gg k_BT/l_p$), we can neglect transverse fluctuations without significantly altering the equilibrium or kinetic properties of the GRM.  The refolding time, $\tau_F^m$ of the RNA mimic (Fig. 3a), which is the WF closure time  \cite{FixmanJCP74II}, can be determined by numerically diagonalizing the Rouse-like matrix with elements \cite{Edwardsbook} ${\mathbf{M}}_{ij}=\frac{1}{2}\delta^2 H/\delta\rv_i\delta\rv_j$.
\\

{\bf Acknowledgements :} We are grateful to Arthur Laporta and N. Toan for useful discussions. This work was supported in part by a grant from the National Science Foundation (CHE 05-14056).


%

\clearpage
\clearpage
\begin{figure}[ht]
\includegraphics[width=3.80in]{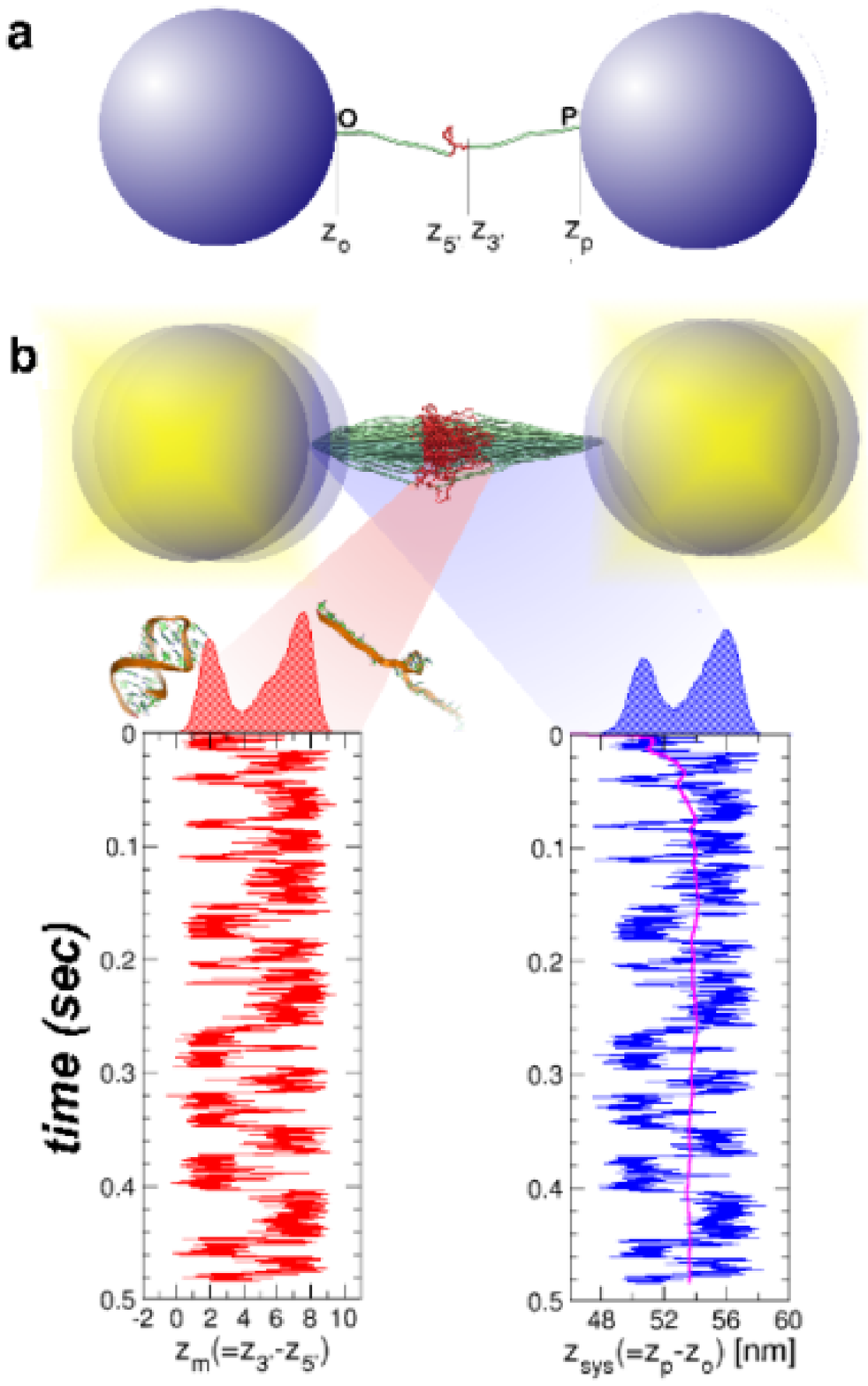}
\caption{\label{Fig1.fig} A schematic diagram of the optical tweezers setup used to measure the hairpin's folding landscape. 
{\bf a}. Two RNA/DNA hybrid linkers are attached to the 5' and 3' ends of the RNA hairpin, and a constant force is applied to one end through the bead.  
{\bf b}. Ensemble of sampled conformations of the H-RNA-H system during the hopping transitions obtained using L=25 nm and $l_p=70$ nm.  
The illustration is created using the simulated structures collected every 0.5 ms.  An example of the time trace of each component of the system, at $f=15.4$ pN, is given $L$ for both linkers is 25 nm.  
$z_m(=z_{5'}-z_{3'})$ and   
$z_{sys}(=z_p-z_o)$ measure the extension dynamics of the 
RNA hairpin 
and of the entire system  
respectively.  
The time averaged value $\overline{z}_T(t)=\frac{1}{t}\int^t_0d\tau z_{sys}(\tau)$ for the time trace of $z_{sys}$ is shown as the bold line.  
The histograms of the extension are shown on top of each column.  
}
\end{figure}
\clearpage
\begin{figure}[ht]
\includegraphics[width=4.50in]{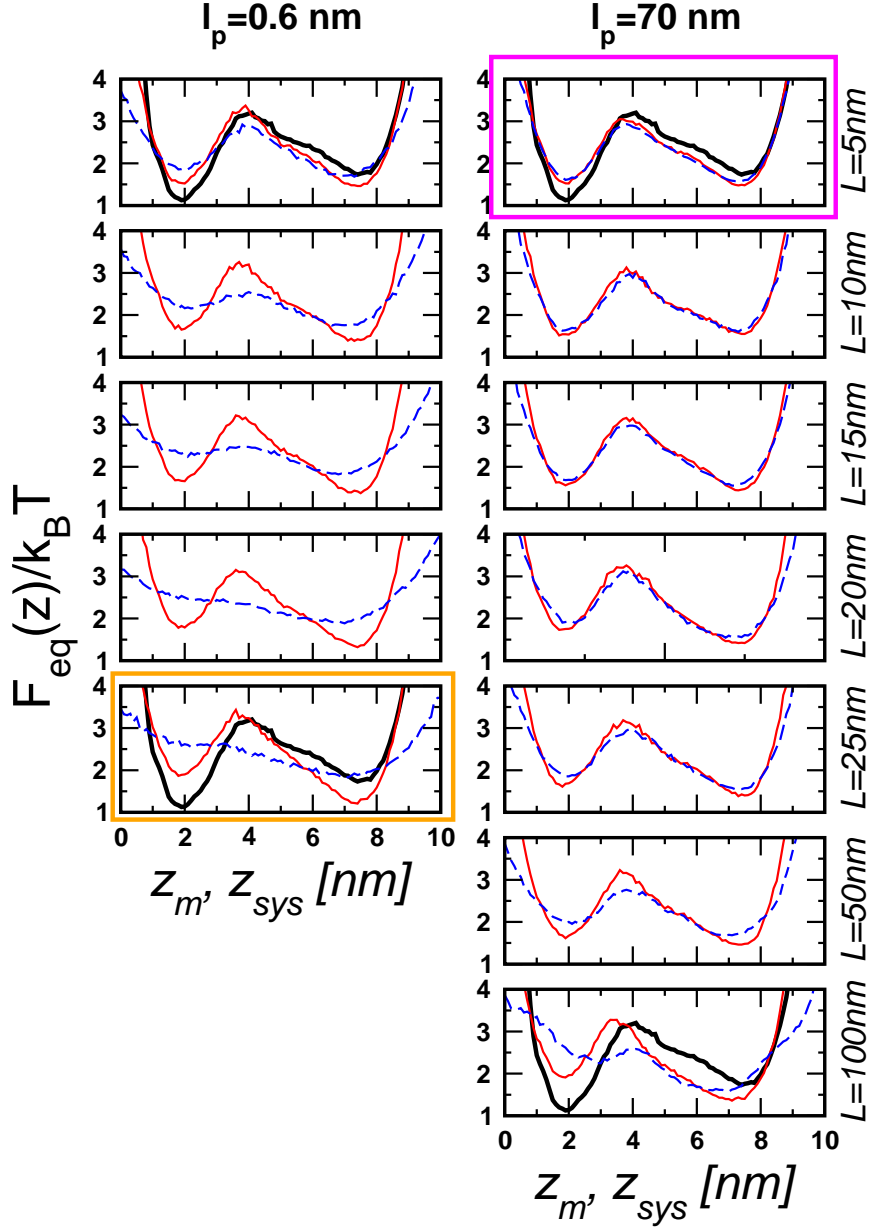}
\caption{\label{Fig2.fig} The free energy profiles, $F_{eq}(z_{sys})$ (dashed line in blue) and  
$F_{eq}(z_m)$ (solid line in red), calculated using the histograms obtained from the time traces $z_{sys}(t)$ and $z_m(t)$ for varying $L$ and $l_p$.  
$F_{eq}(z_{sys})$ and $F_{eq}(z_m)$ for a given $l_p$ and different $L$ are plotted in the same graph to highlight the differences.  
The  intrinsic free energy $F_{eq}^o(z_m)$, the free energy profile in the \emph{absence} of handle, is shown in black.  The condition that produces the least deviation ($l_p=70$ nm, $L=5$ nm) and the condition of maximal difference ($l_p=0.6$ nm, $L=25$ nm) between $F_{eq}(z_m)$ and $F_{eq}(z_{sys})$ are enclosed in the magenta and orange boxes, respectively.  
}
\end{figure}
\clearpage
\begin{figure}[ht]
\includegraphics[width=2.50in]{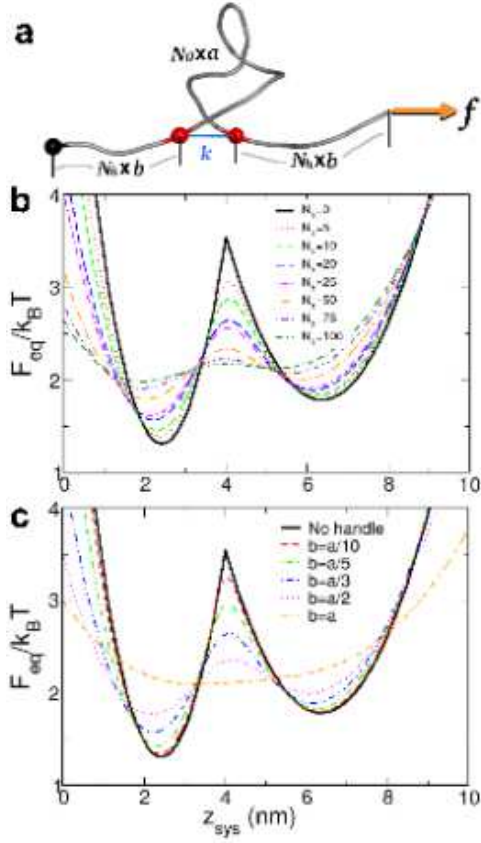}
\caption{\label{Fig3.fig} Free energy profiles for the GRM.  {\bf a}.  A schematic diagram of the GRM, showing the number of monomers ($N_0$ and $N_h$) and Kuhn lengths ($a$ and $b$) in each region of the chain, the harmonic interaction between the ends of the RNA mimic, and the external tension.  {\bf b}. The free energy profile for a fixed $b(=a/3)$ and increasing $N_h$ as a function of $R_{sys}\approx z_{sys}$.  The barrier heights decrease and the well depths increase for increasing $N_h$.  {\bf c}:  The free energy profile for fixed $N_h=20$ and varying $b$. The barrier heights decrease and the well depths increase for increasing $b$.  In both b and c, the profiles are shifted so that the positions of the local maxima and minima coincide with those of the intrinsic free energy (with $N_h=0$).}
\end{figure}
\clearpage
\begin{figure}[ht]
\includegraphics[width=7.0in]{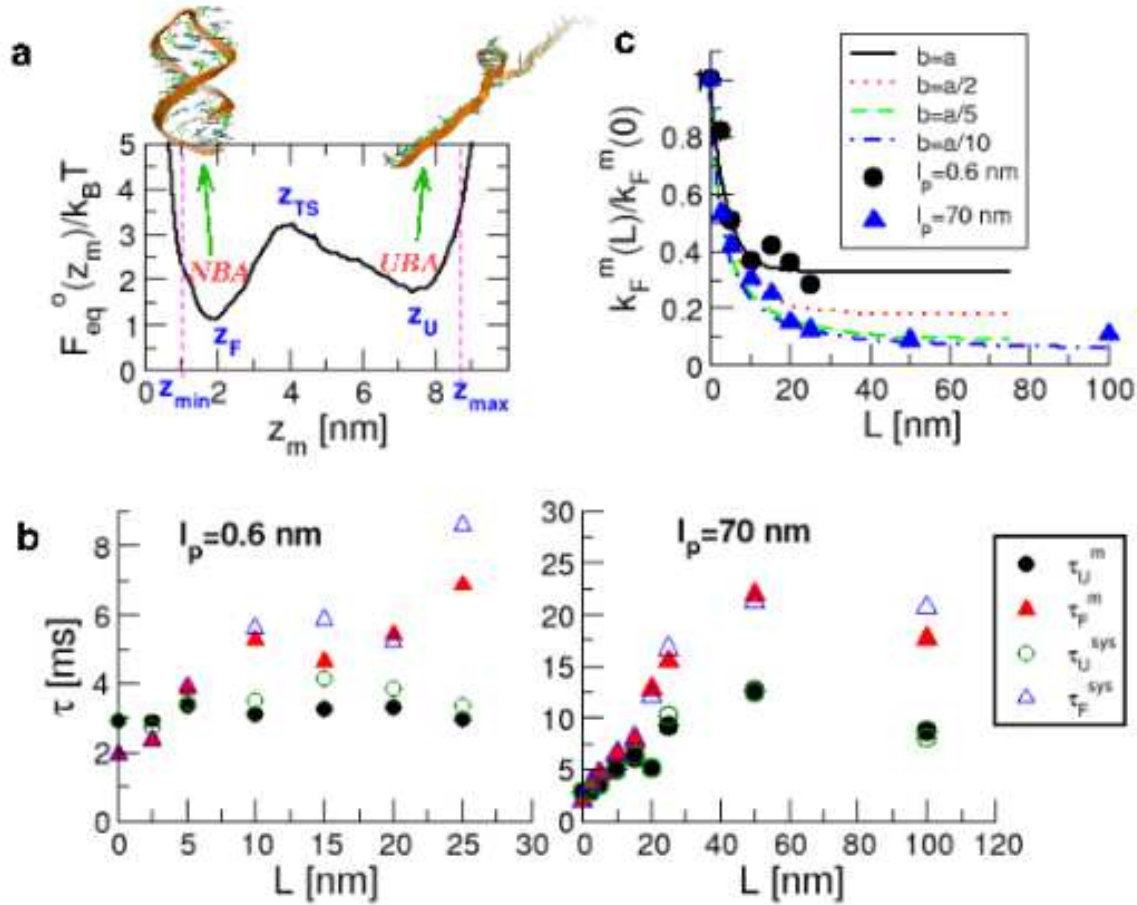}
\caption{\label{Fig4.fig} {\bf a}. The free energy profile for P5GA with $L=0$ nm.
{\bf b}. The transition times at $f=f_m$, obtained using $z_m(t)$ (filled symbols) and $z_{sys}(t)$ (empty symbols).
 The ratio $\tau^m_{U(F)}/\tau^{sys}_{U(F)}\approx 1$, which shows that $z_{sys}(t)$ mirrors the hopping of P5GA.
{\bf c}. Folding rate $k_F^m(L)/k_F^m(0)$ as a function of $L$ for varying $b$, using the GRM.  The plots show $b/a=$1, 1/2, 1/5, and 1/10.  The simulation results for P5GA are also shown as symbols, to emphasize that the GRM accounts for the hopping kinetics in the H-RNA-H system accurately.
}
\end{figure}
\clearpage
\begin{figure}[ht]
\includegraphics[width=2.50in]{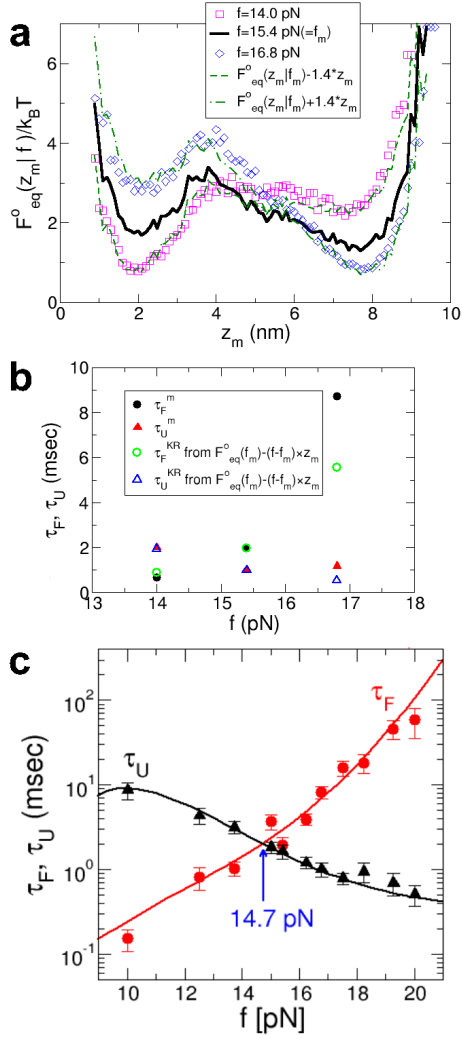}
\caption{\label{fig5.fig} {\bf a}. Comparison of the measured free energy profiles (symbols) with the shifted free energy profiles $\beta F^o_{eq}(z_m|f_m)-\beta(f-f_m)\cdot z_m$.  {\bf b}. Folding and unfolding times as a function of force $f=14$ pN $<f_m$, $f=15.4$ pN $\approx f_m$, and $f=16.8$ pN $>f_m$.  $\tau_{F(U)}^m$ is obtained from the time trace in Figure 2B in Ref \cite{HyeonBJ07} at each force, while $\tau_{F(U)}^m$ is computed using the tilted profile $\beta F_{eq}^o(z_m|f)=\beta F_{eq}^o(z_m|f_m)-\beta(f-f_m)\cdot z_m$ in equation (\ref{eqn:MFPT}).  {\bf c}. Folding and unfolding times using the GRM.  Symbols are a direct simulation of the GRM (error bars are standard deviation of the mean).  The solid lines are obtained using the Kramers theory (equation (\ref{eqn:MFPT})).  We choose $D_U\approx 3D_0$, so that that the simulated and Kramers times agree at $f=f_m$.   The position of each basin of attraction as a function of force for the GRM is given by $z_{U}\approx N_0a^2 \beta f/3$ and $z_{F}\approx N_0a^2 \beta f/(3 +2N_0a^2\beta k)$.}
\end{figure}
\clearpage

{\bf SUPPORTING INFORMATION (SI)}\\

{\bf Fluctuations of the handle polymer under tension :}
Using the force clamp simulations of the RNA hairpin in the presence of handles,
the dynamics of the fluctuations of the handle can be independently extracted by probing the time-dependent changes in the 5' and 3' ends of the RNA molecule.
The distribution of the longitudinal fluctuations ($z_H^{5'}\equiv z_{5'}-z_o$ and $z_H^{3'}\equiv z_p-z_{3'}$) and the dispersion in the transverse fluctuations ($x_{5'}$ or $y_{5'}$ and $x_{3'}$ or $y_{3'}$)
are shown in SI Fig. 6.
Assuming that the force ${\mathbf{f}}=f_{\parallel}\hat{\mathbf{e}}_{\parallel}+f_{\perp}\hat{\mathbf{e}}_{\perp}$ is decomposed into $f=|{\mathbf{f}}|\approx f_{\parallel}+f_{\perp}^2/2f_{\parallel}+\mathcal{O}(f^4_{\perp})$ (with $f_{\parallel}\gg f_{\perp}$), and using the partition function of the worm-like chain polymer under tension, $Z=\int d\Omega e^{-(H_{WLC}/k_BT-\vec{f}\cdot\vec{R}/k_BT)}$ \cite{Marko05PRE},
one can express the longitudinal fluctuation as
\begin{equation}
\langle\delta R_{\parallel}^2\rangle=k_BT\frac{d\langle R_{\parallel}\rangle}{df}\sim \left\{ \begin{array}{ll}
     Ll_p & \mbox{for $f \lesssim \frac{k_BT}{l_p}$}\\
     L \,l_p^{-1/2}(f/k_BT)^{-3/2}& \mbox{for $f\gg \frac{k_BT}{l_p}$}\end{array}\right. ,
\label{eqn:longitudinal}
\end{equation}
where the force extension relations of a worm-like chain $R_{\parallel}/L\approx fl_p/k_BT$ for $f<k_BT/l_p$ and $R_{\parallel}/L\approx 1-\sqrt{k_BT/4l_pf}$ for $f>k_BT/l_p$ are used for small and large forces, respectively \cite{Marko95Macro}.
These results are consistent with the fluctuations observed in the simulations.  When $f<k_BT/l_p$, the transverse fluctuations are independent of the force, and are determined solely by the nature of the linker.  For $f=f_m\approx 15.4$ pN, the tension is in the regime that satisfies $f>k_BT/l_p$ for both values of $l_p$ used in the simulations, and the longitudinal fluctuations $\langle\delta R_{||}^2\rangle$ decrease as the stiffness of the polymer increases for all $L$ (SI Fig.6a). The distribution of the extensions coincide for both the 3' and 5' ends of the handles for all $L$ and $l_p$ (SI Fig. 6a).
This suggests that the constant force applied at the point $z_p$ propagates uniformly throughout the whole system.

The transverse fluctuations are given by
\begin{equation}
\langle\delta R_{\perp}^2\rangle=(k_BT)^2\frac{\partial^2\log{Z}}{\partial f_{\perp}^2}|_{f_{\parallel}=f,f_{\perp}=0}=\frac{k_BT}{f}\langle R_{\parallel}\rangle\approx \left\{ \begin{array}{ll}
   Ll_p& \mbox{for $f\lesssim \frac{k_BT}{l_p}$}\\
   \frac{Lk_BT}{f}\left(1-\frac{1}{2}\sqrt{\frac{k_BT}{l_pf}}\right)& \mbox{for  $f\gg \frac{k_BT}{l_p}$}\end{array}\right. .
\label{eqn:transverse}
\end{equation}
The transverse fluctuations also decrease as $f(>k_BT/l_p)$ increases, with a different power.  It is worth noting that the transverse fluctuations, which also increase as $L$ increases, are nearly independent of the handle stiffness if $f>k_BT/l_p$.  The standard deviations of distributions are plotted with respect to the contour length at each bending rigidity (SI Fig. 6b).
The fit shows that $\sigma\sim 0.28\times L^{1/2}$ nm and $\sigma\sim 0.30\times L^{1/2}$ nm for $l_p=0.6$ nm and $l_p=70$ nm, respectively, which is consistent with the analysis in equation \ref{eqn:transverse}.\\
{\bf Determination of the persistence length of the handles :} In order to determine the persistence length of the handles, we numerically generated the end-to-end distribution function $P(R)$ of the free handles in the absence of tension, and fit the simulated distribution to the analytical result
\cite{HaBook,HyeonJCP06}
\begin{equation}
P_{WLC}(R)=\frac{4\pi C\rho^2}{L(1-\rho^2)^{9/2}}\exp{\left(-\frac{\alpha}{1-\rho^2}\right )}
\end{equation}
with $\rho=R/L$ and $\alpha=3L/4l_p$.  The normalization constant $C=$$\left[\pi^{3/2}e^{-\alpha}\alpha^{3/2}(1+3\alpha^{-1}+15\alpha^{-2}/4)\right]^{-1}$ ensures $\int_0^LdR\ P_{WLC}(R)=1$.
\\

{\bf Quantifying the synchronization of $z_{sys}(t)$ and $z_m(t)$ : }
The origin of the small discrepancy between $F_{eq}(z_{sys})$ and $F_{eq}(z_m)$ when the handles are flexible can be found by comparing $z_{sys}(t)$ with $z_m(t)$ for the two extreme cases in Fig. 2 of the main text.
To quantitatively express the synchronization between $z_{sys}(t)$ and $z_m(t)$, we defined a correlation function at each time $t$ using
\begin{equation}
C(t)=\frac{z_{sys}(t)-z_{sys}^{TS}}{z_{sys}^{TS}}\times\frac{z_m(t)-z_m^{TS}}{z_m^{TS}} 
\end{equation}
where $z_{sys}^{TS}$ and $z_m^{TS}$ are the positions of the transition states determined from $F_{eq}(z_{sys})$ and $F_{eq}(z_m)$ respectively.  If $C(t)>0$ at time $t$, both $z_{sys}(t)$ and $z_m(t)$ are in the same basins of attraction, i.e. the status of $z_m(t)$ is correctly detected by the
measurement through the handles. If $C(t)<0$, then the information of $z_m(t)$ is lost due to fluctuations or the slow response of the handles.  The near perfect synchronization between $z_{sys}(t)$ and $z_m(t)$ for $l_p=60$ nm and $L=5$ nm  are reflected in $C(t)>0$ for almost all $t$.  Thus, when the handles are stiff, $z_m(t)\approx z_{sys}(t)-2L$, which implies that $z_{sys}(t)$ faithfully reflects the dynamics ($z_m(t)$) of the hairpin.  In contrast, with $l_p=0.6$ nm and $L=25$ nm, $z_{m}(t)$ can not be determined from $z_{sys}(t)$ using $z_{sys}(t)-2\times L \neq z_m(t)$.
The amplitudes of $z_{sys}(t)$ are typically larger than that of $z_m(t)$, leading to $C(t)<0$ occasionally (shown by an arrow on the right plot in SI Fig. 7).
The histograms of $P(C)$ for the two extreme cases show that the dynamics between $z_{sys}(t)$ and $z_m(t)$ are more synchronous for the rigid and short handles ($0.0<P(C)<0.5$) than for the flexible and longer handles ($-0.05<P(C)<0.2$) (see the graph at the bottom in the SI Fig.7).  The finding that short and stiff ($L/l_p\sim O(1)$) handles minimize the differences between $F_{eq}(z_m)$ and $F_{eq}(z_{sys})$ is related to the tension-dependent fluctuations in the linkers.\\

\clearpage
\begin{figure}[ht]
\includegraphics[width=6.00in]{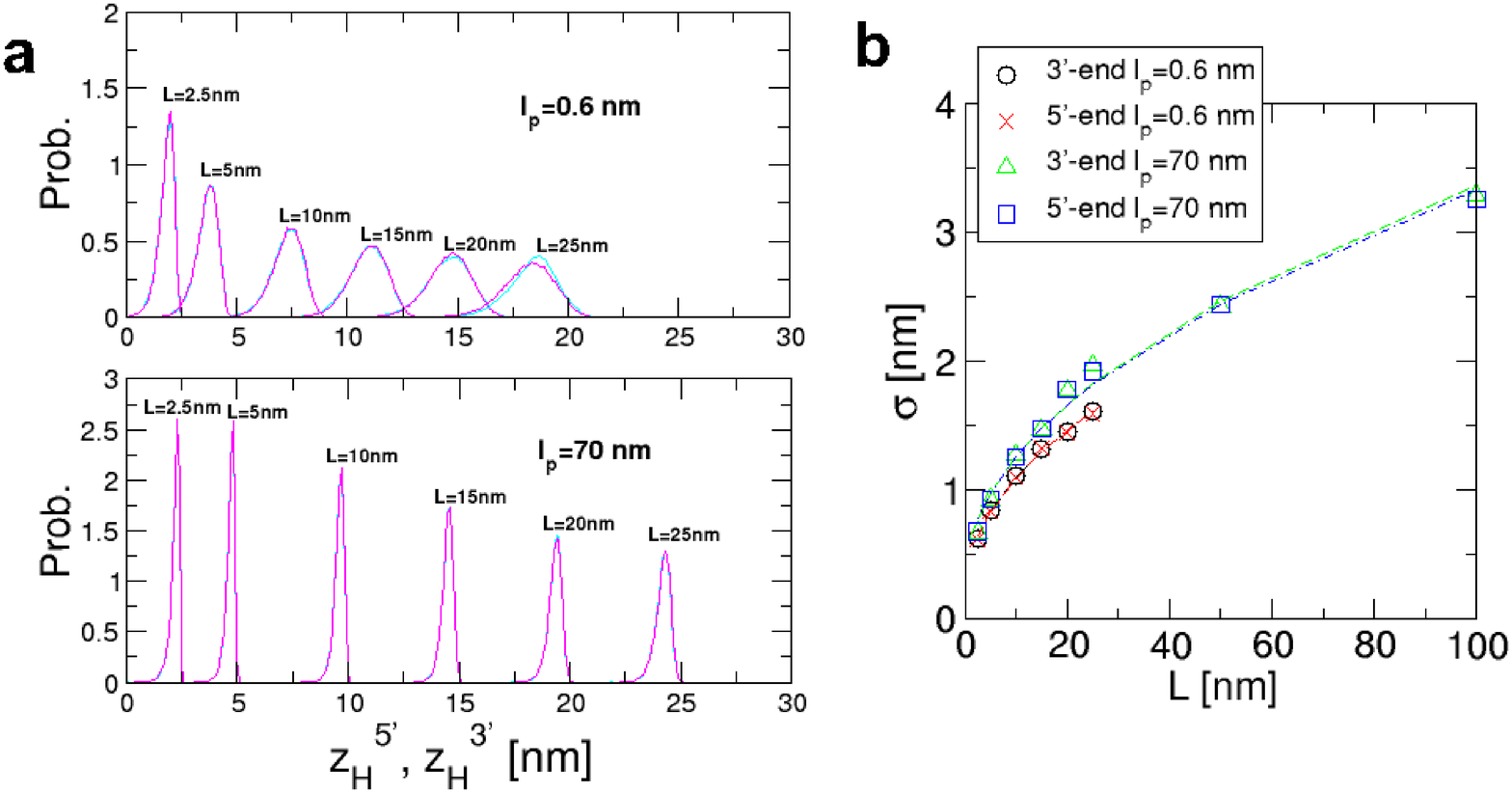}
\caption{\label{SI_Fig1.fig} Fluctuations of the handles with varying lengths and flexibilities at $f=15.4$ pN.
{\bf a}. Longitudinal fluctuations of the handle attached at the 5' and 3' sides of the RNA hairpin.
{\bf b}. Transverse fluctuations are fit to a Gaussian distribution, and the standard deviation ($\sigma$) is plotted as a function of the contour length and flexibility.
}
\end{figure}
\begin{figure}[ht]
\includegraphics[width=5.00in]{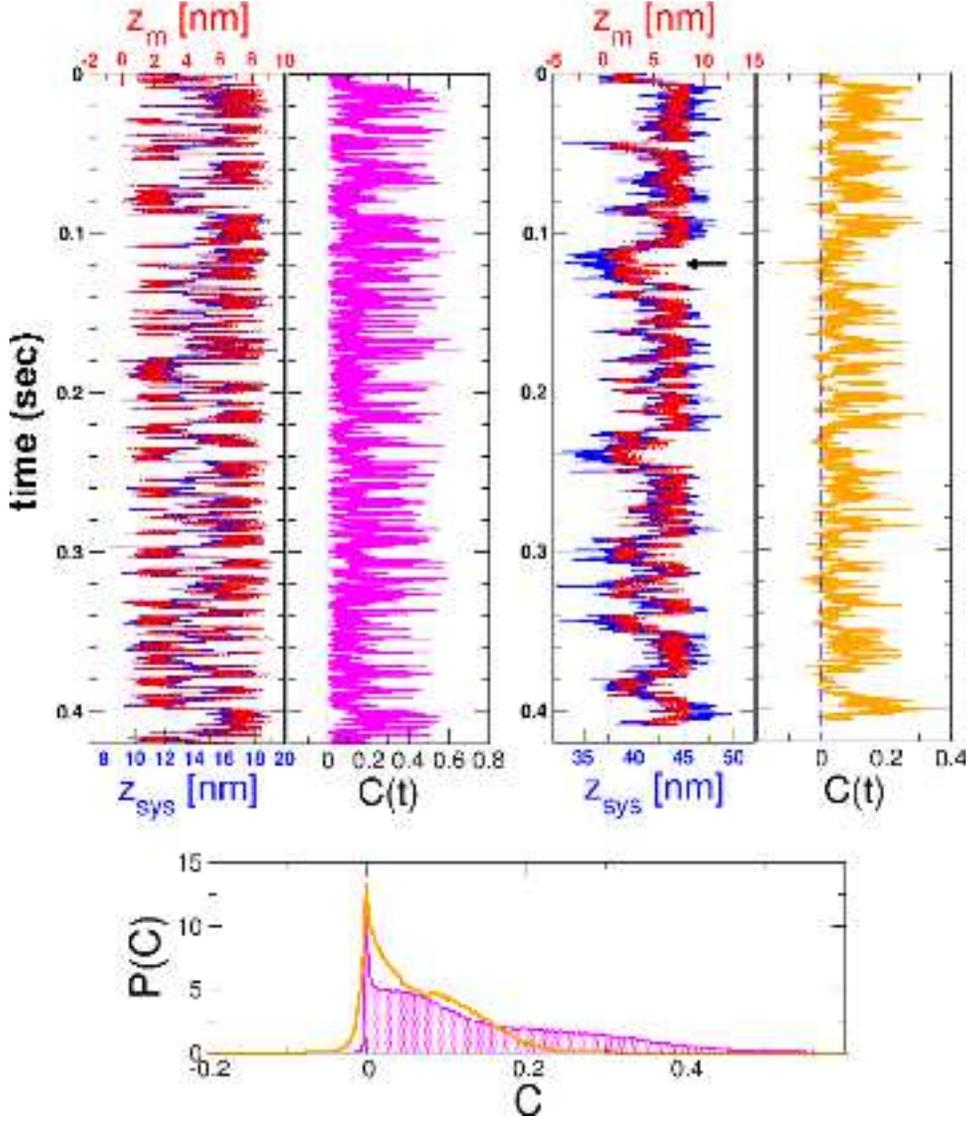}
\caption{\label{trace.fig} The time traces, $z_{sys}(t)$ and $z_m(t)$, for two extreme cases that produce the free energy profiles in the magenta and orange boxes in Fig. 2 of the main text, are overlapped to show the differences.
For $l_p=70$ nm and $L=5$ nm, both the phase and amplitude between $z_{sys}(t)$ and $z_m(t)$ coincide throughout the time series, while for $l_p=0.6$ nm and $L=25$ nm the amplitude of $z_{sys}(t)$ are larger than $z_m(t)$ and the phase between
$z_{sys}(t)$ and $z_m(t)$ is occasionally offset from one another.
The correlation measure $C(t)$ quantifies the synchrony between $z_{sys}(t)$ and $z_m(t)$ at time $t$.
The histograms of $C(t)$ show that the time trace for $l_p=$70 nm and $L=5$ nm is
more synchronized than the one for $l_p=0.6$ nm and $L=25$ nm.
}
\end{figure}

\begin{figure}[ht]
\includegraphics[width=6.00in]{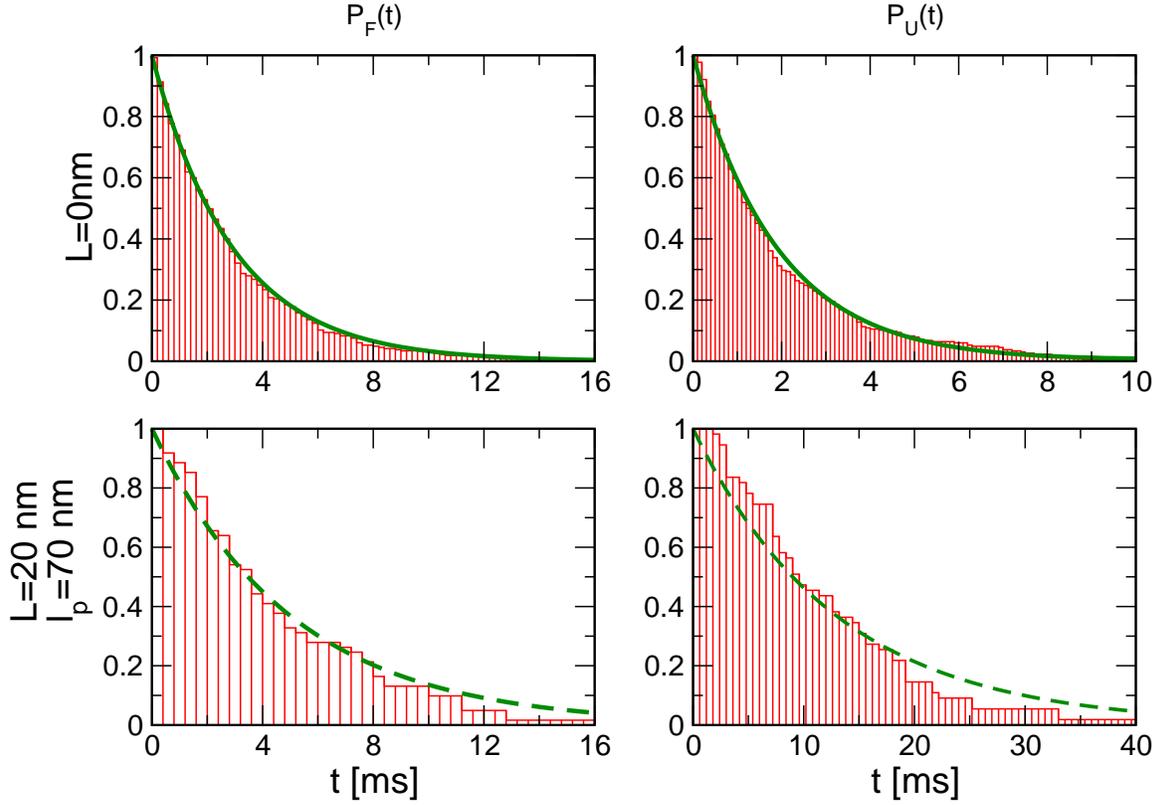}
\caption{\label{Survival.fig} The survival probabilities $P_F(t)$ and $P_U(t)$ are fit to a single exponential function to calculate $\tau_U$ and $\tau_F$. For $L$=0 nm,
$\tau_U=2.9$ ms and $\tau_F=1.9$ ms. For $L$=20 nm, $\tau_U=5.0$ ms and $\tau_F=12.1$ ms. The quality of the fits for $L=20$ nm (dashed lines) is not as good as for $L=0$ nm. The survival probabilities show lag phases for both unfolding and refolding.}
\end{figure}

\begin{figure}[ht]
\includegraphics[width=4.00in]{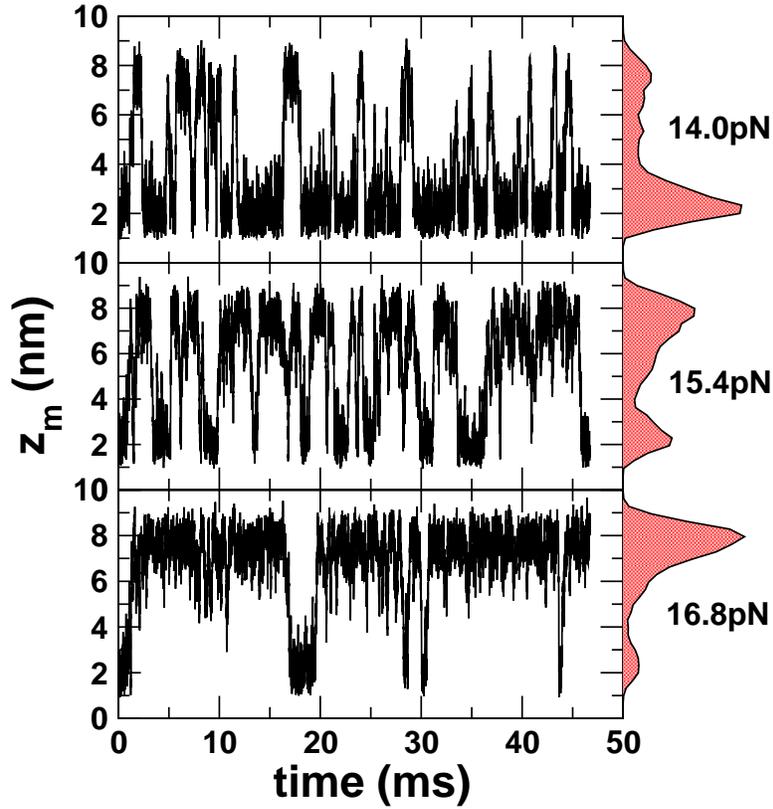}
\caption{\label{SI4.fig} Time traces of molecular extension under tension $f=$14.0, 15.4, and 16.8 pN, and corresponding distribution $P(z_m|f)$ at each force. The distributions are converted to the free energy profile in Fig.5a by using $F_{eq}^o(z_m)/k_BT=-\log{P(z_m)}$.  Note that the hairpin is pinned in the UBA at $f=16.8$pN ($>f_m$) with infrequent transitions to the NBA.  Just as in experiments \cite{Block06Science}, accurate measurement of $F_{eq}(z_m)$ is possible only at $f\approx f_m$, where multiple hopping events between the NBA and UBA can be observed.
}
\end{figure}

\end{document}